\documentclass[11pt,twoside,letterpaper]{article} 
\usepackage{amsmath,amssymb}%
\usepackage{bm}%
\usepackage{epsfig}
\usepackage{times,fancyhdr}
\usepackage[english]{babel}%
\usepackage{url}
\usepackage{amsmath,amsthm,amscd,amssymb}
\usepackage{latexsym}
\usepackage{graphicx,epsfig}
\RequirePackage{natbib}%

\def\tablename{Table}
\makeatletter
\renewcommand{\fnum@table}[1]{\bf\tablename~\thetable.}
\makeatother

\def\rfr#1{eq. (\ref{#1})}
\def\rfrs#1#2{eq. (\ref{#1})-eq. (\ref{#2})}

\def\Po{P_{\rm b}}
\def\nkep{n}
\def\Pkep{P_{\rm b}}

\def\dert#1#2{\frac{{{d}}{#1}}{{{d}}{#2}}}              

\def\bar{\begin{eqnarray}}
\def\ear{\end{eqnarray}}
\def\bb{\bibitem}
\def\eqi{\begin{equation}}
\def\eqf{\end{equation}}
\def\eqia{\begin{eqnarray}}
\def\eqfa{\end{eqnarray}}
\def\rp#1#2{{#1\over#2}}

\def\lb#1{\label{#1}}

\def\olt{\dot\omega_{\rm GM}}
\def\oge{\dot\omega_{\rm 1PN}}
\def\ogge{\dot\omega_{\rm 2PN}}
\def\oms{\dot\omega_{\rm meas}}

\def\psr{PSR J0737-3039A/B\ }

\def\cen{1-e^2}

\def\oc2{$\mathcal{O}(c^{-2})$}
\def\opn{\dot\omega_{\rm 1PN}}
\def\Pop{\rp{\Po}{2\pi}}
\def\Ts{T_{\odot}}
\def\xa{x_{\rm A}}
\def\xb{x_{\rm B}}

\def\bds#1{\boldsymbol{#1}}

\usepackage{url}
\usepackage{amsmath,amsthm,amscd,amssymb}
\usepackage{latexsym}
\usepackage{graphicx,epsfig}

\RequirePackage{color}

\begin{document}

\title{
{\begin{flushleft}
\vskip 0.45in
{\normalsize\bfseries\textit{Chapter~4}}
\end{flushleft}
\vskip 0.45in
\bfseries\scshape{Some Applications of Binary Pulsars to Fundamental Physics}}}
 \author{\bfseries{\itshape  Lorenzo Iorio \thanks{Email address: lorenzo.iorio@libero.it}}\\
F.R.A.S. Permanent address for \\
correspondence: Viale Unit\`{a} di Italia 68, 70125, Bari (BA), Italy.}

 \date{}
\maketitle
\thispagestyle{empty}
\setcounter{page}{1}

\thispagestyle{fancy}
\fancyhead{}
\fancyhead[L]{In: Pulsars: Theory, Categories and Applications  \\ 
Editor: Alexander D. Morozov} 
\fancyhead[R]{ISBN:  978-1-61668-919-3  \\
\copyright~2010 Nova Science Publishers, Inc.}
\fancyfoot{}
\renewcommand{\headrulewidth}{0pt}

\begin{abstract}
Binary systems containing at least one radiopulsar are excellent laboratories to test several aspects of fundamental physics like matter properties in conditions of extreme density and theories of gravitation like the Einstein's General Theory of Gravitation (GTR) along with modifications/extensions of it.
In this Chapter we focus on the perspectives on measuring the moment of inertia of the double pulsar, its usefulness in testing some modified models of gravity, and the possibility of using the mean anomaly as a further post-Keplerian orbital parameter to probe GTR.

\vspace*{11pt}

\vspace*{11pt}
\end{abstract}

\section{Introduction}
In this Chapter we discuss some applications of astrophysical binary systems containing at least one radiopulsar to fundamental physics.
In particular, in Section \ref{inertia} we investigate the perspectives on measuring the pulsar's moment of inertia  \citep{Ioruno,Kranew} through the extension of the general relativistic Lense-Thirring effect to the double pulsar system J0737-3039. Section \ref{modified} shows how the double pulsar can be used to put on the test some models of modified gravity \citep{Iordue}. In Section \ref{meananomaly} we deal with the possibility of using the mean anomaly as a further post-Keplerian parameter useful to test the gravitolectric part of the General Theory of  Relativity (GTR) \citep{Iortre}; for a proposal concerning a new post-Keplerian parameter to test different aspects of gravitomagnetism with respect to the Lense-Thirring effect, see \citep{rugtart}.
\section{The Moment of Inertia of Pulsar J0737-3039A: Prospects for its Measurement}\lb{inertia}
 The  measurement of the moment of inertia $I$ of a neutron star at a $10\%$ level of accuracy or better would be of crucial importance for effectively constraining the Equation-Of-State (EOS) describing  matter inside neutron stars \citep{Mor04,Bej05,Lat05,Lav07}.

 After the discovery of the double pulsar \psr system \citep{Bur03}, whose relevant orbital parameters  are listed in Table \ref{tavola}, it was often argued that such a measurement for the A pulsar via the post-Newtonian gravitomagnetic spin-orbit periastron precession \citep{Bar75a,Dam88,Wex95} would be possible after  some years of accurate and continuous timing.
 \begin{table}[!h]

\caption{ Relevant Keplerian and post-Keplerian parameters of the binary system
\psr \citep{Kra06}. The orbital
period $P_{\rm b}$ is measured with a precision of $4\times 10^{-6}$ s. The projected semimajor axis is
defined as $x\doteq(a_{\rm bc}/c)\sin i,$  where $a_{\rm bc}$ is the
barycentric semimajor axis (the relative semimajor axis $a = (x_{\rm A} + x_{\rm B})c/\sin i$), $c$ is the speed of light and $i$ is the angle
between the plane of the sky, perpendicular to the line-of-sight,
and the orbital plane. The eccentricity is $e$. The best determined post-Keplerian parameter is, to date, the periastron rate $\dot\omega$ of the component A. The phenomenologically determined post-Keplerian parameter $s$, related to the general relativistic Shapiro time delay,
is equal to $\sin i$; we  have
conservatively quoted the largest error in $s$ reported in
\citep{Kra06}.  The other post-Keplerian parameter related to the Shapiro delay, which is used in the text, is $r$.
}
\label{tavola}
\begin{center}
{\scriptsize{\begin{tabular}{lllllll}

\noalign{\hrule height 1.5pt}

 $P_{\rm b}$ (d)& $x_{\rm A}$ (s) & $x_{\rm B}$ (s) & $e$ & $\dot\omega$ (deg yr$^{-1}$) & $s$ & $r$ ($\mu$s)\\
\hline
$0.10225156248(5)$ & $1.415032(1)$ & $1.5161(16)$ &  $0.0877775(9)$ & $16.89947(68)$  & $0.99974(39)$  &  $6.21(33)$ \\

\hline

\noalign{\hrule height 1.5pt}

\end{tabular}}}
\end{center}
\end{table}
%
%
%
%
%
\citet{Lyn04} write: ``Deviations
from the value predicted by general relativity may be caused by contributions from spin-orbit
coupling \citep{Bar75b}, which is about an order of magnitude larger than for PSR B1913+16. This
potentially will allow us to measure the moment of inertia of a neutron star for the first time \citep{Dam88, Wex95}.''

\pagestyle{fancy}
\fancyhead{}
\fancyhead[EC]{Lorenzo Iorio}
\fancyhead[EL,OR]{\thepage}
\fancyhead[OC]{Some Applications of Binary Pulsars to Fundamental Physics}		
\fancyfoot{}
\renewcommand\headrulewidth{0.5pt}

According to \citet{Lat05}, ``measurement of the spin-orbit perihelion advance seems possible.''

In \citep{Kra06} we find:  ``A future determination of the system geometry and the measurement of two other PK parameters at a level of precision similar
to that for $\dot\omega$, would allow us to measure the moment of inertia of a neutron star for the
first time \citep{Dam88, Wex95}.  [...] this measurement is potentially very difficult [...] The double pulsar [...] would also give insight into the nature of super-dense matter.''

In \citep{Dam07} it is written: ``It was then pointed
out \citep{Dam88} that this gives, in principle, and indirect way of measuring the moment
of inertia of neutron stars [...]. However, this can be done only if one measures,
besides\footnote{The parameter $k$ is directly related to the periastron precession $\dot\omega$.} $k$, two other PK parameters with $10^{-5}$ accuracy. A rather tall order
which will be a challenge to meet.''

Some more details are released by \citet{Kra05}: ``[...] a potential measurement
of this effect allows the moment of inertia of a neutron
star to be determined for the first time \citep{Dam88}. If two parameters, e.g. the Shapiro parameter $s$ and
the mass ratio $\mathcal{R}$, can be measured sufficiently accurate,
an expected $\dot\omega_{\rm exp}$ can be computed from the intersection
point.''

Here we will examine with some more details the conditions which would make feasible to measure $I_{\rm A}$ at $10\%$ or better in the \psr system in view of the latest timing results \citep{Kra06}.
In particular, we will show how important the impact of the mismodelling in the known precessional effects affecting the periastron rate of \psr is
if other effects must be extracted from such a post-Keplerian parameter.  Such an analysis will turn out to be useful also for purposes other than measuring gravitomagnetism like, e.g., putting more severe and realistic constraints on the parameters entering various models of modified gravity (See Section \ref{modified}). Indeed, in doing that for, e.g., a uniform cosmological constant $\Lambda$ \citet{Jet} took into account only the least-square covariance sigma of the estimated periastron rate ($6.8\times 10^{-4}$ deg yr$^{-1}$): the systematic bias due to  the first post-Newtonian (1PN) periastron precession was neglected. Concerning the use of the mean anomaly $\mathcal{M}$ for testing 1PN effects on it in pulsar systems, see Section \ref{meananomaly}.
\subsection{The Systematic Uncertainty in the 1PN Periastron Precession}\lb{kazzam}
By assuming $I\approx 10^{38}$ kg m$^2$ \citep{Mor04, Bej05}, the gravitomagnetic spin-orbit periastron precession is about $\olt\approx 10^{-4}$ deg yr$^{-1}$, while the error $\delta\oms$ with which the periastron rate is phenomenologically estimated from timing data is currently $6.8\times 10^{-4}$ deg yr$^{-1}$ \citep{Kra06}. In order to measure the gravitomagnetic  effect$-$and, in principle, any other dynamical feature affecting the periastron$-$ $\delta\oms$ is certainly of primary importance, but it is not the only source of error to be carefully considered: indeed,
there are other terms contributing to the periastron precession (first and second post-Newtonian, quadrupole, spin-spin  \citep{Bar75a, Dam88, Wex95}) which must be subtracted from  $\oms$, thus  introducing a further systematic uncertainty due to the propagation of the errors in the system's parameters entering their analytical expressions. A preliminary analysis of such aspects, can be found in \citep{Lat05}. However, apart from the fact that its authors make use of the value for $i$ measured with the scintillation method \citep{Col05} which is highly uncertain for the reasons given below, in using the third Kepler law to determine the sum of the masses they also confound the relative projected semimajor axis\footnote{Our $a$ must not be confused with the one used in \citep{Lat05} which is, in fact, $a_{\rm bc}$. } $a\sin i$ (see \rfr{ohe}) with the barycentric projected semimajor axis $x$, which is the true measurable quantity from timing data, so that their analysis cannot be
considered reliable.  The semimajor axis $a$ of the relative motion of A with respect to B  in a binary system is just the sum of the semimajor axes $a_{\rm bc}$ of A and B with respect to the system's barycenter, i.e.
$a = a_{\rm bc}^{\rm A} + a_{\rm bc}^{\rm B}$.

Let us, now, consider the largest contribution to the periastron rate, i.e. the 1PN precession \citep{Dam86, Dam92}
\eqi \oge = \rp{3}{(1-e^2)}\left(\rp{P_{\rm b}}{2\pi}\right)^{-5/3}(T_{\odot}M)^{2/3},\lb{oge}\eqf
where $T_{\odot}=G{\mathfrak{M}}_{\odot}/c^3$ and $M=m_{\rm A}+m_{\rm B}$, in units of Solar mass $\mathfrak{M}_{\odot}$. It is
often referred to as gravitoelectric in the weak-field and slow-motion approximation: in the context of the Solar System it is the well known Einstein Mercury's perihelion precession of about 43 arcsec cy$^{-1}$. Thus,
\begin{equation}\left\{\begin{array}{lll}
\olt = \oms-\oge-\dot\omega_{\rm 2PN},\\\\
\delta\olt \leq \delta\oms + \delta\oge +\delta\dot\omega_{\rm 2PN}
 \lb{prece}
\end{array}\right.\end{equation}
The sum of the masses $M$  enters \rfr{oge}; as we will see, this implies that  the relative semimajor axis $a$ is required as well.
For consistency reasons,  the values of such parameters used to calculate \rfr{oge} should have been obtained independently of the periastron rate itself. We will show that, in the case of PSR J0737-3039A/B, it is possible.

Let us start from the relative semimajor axis
\eqi a=
  \left(1+\mathcal{R}\right)\left(\rp{cx_{\rm A}}{\sin i}\right)= 8.78949386\times 10^8\ {\rm m}\lb{ohe}.\eqf
It is built in terms of $\mathcal{R}$, the projected semimajor axis $x_{\rm A}$ and $\sin i$;  the phenomenologically estimated
post-Keplerian parameter $s$ determining the shape of the logarithmic Shapiro time delay can be identified with $\sin i$ in GTR and  the
ratio $\mathcal{R}\doteq x_{\rm B}/x_{\rm A}$ has been obtained from the phenomenologically determined projected semimajor axes, being equal to the ratio of the masses in any Lorentz-invariant theory of gravity  \citep{Dam85, Dam88, Dam92} \eqi \mathcal{R}=\rp{\mathfrak{m}_{\rm A}}{\mathfrak{m}_{\rm B}}+\mathcal{O}\left(\rp{v^4}{c^4}\right).\eqf
The uncertainty in $a$ can be conservatively evaluated as
\eqi \delta a \leq   \delta
a|_\mathcal{R} + \delta a|_s + \delta a|_{x_{\rm A}} ,\eqf with
\begin{equation}\left\{\begin{array}{lll}
\delta a|_\mathcal{R}\leq  \left(\rp{cx_{\rm A}}{s}\right)\delta \mathcal{R} = 466,758\ {\rm m},\\\\
\delta a|_s\leq a\left(\rp{\delta s} s\right) = 342,879\ {\rm m},\\\\
\delta a|_{x_{\rm A}} \leq a\left(\rp{\delta x_{\rm A}}{x_{\rm A}}\right)=621 \ {\rm m}.

 \lb{erra}
\end{array}\right.\end{equation}
Thus, \eqi \delta a \leq 810,259\ {\rm m}.\lb{longo}\eqf
  \rfr{longo} yields a relative uncertainty of \eqi \rp{\delta a}{a} = 9\times 10^{-4}.\eqf It is
important to note that $x_{\rm B}$, via $\mathcal{R}$, and $s$ have a major impact on the
overall uncertainty in $a$; our estimate has to be considered as
conservative because we adopted for $\delta s$ the largest value
quoted in \citep{Kra06}. In regard to the inclination, we did not use the more precise value for $i$ obtained from scintillation measurements in\footnote{See also \citep{Lyu05}.} \citep{Col05} because it
is inconsistent with that
derived from timing measurements \citep{Kra06}. Moreover, the scintillation method is model-dependent and it is not
only based on a number of assumptions about the interstellar medium, but it is also much
more easily affected by various other effects.   However, we will see that also $x_{\rm A}$ has a non-negligible impact.
Finally, let us note that we purposely linearly summed up the individual sources of errors in view of the  existing correlations among the various estimated parameters \citep{Kra06}.

Let us, now, determine the sum of the masses: recall that it must not come from the periastron rate itself. One possibility is to use the phenomenologically determined orbital period $P_{\rm b}$ and the third Kepler law getting\footnote{In principle, also the 1PN correction to the third Kepler law calculated by \citep{Dam86} should be included, but it does not change the error estimate presented here.}
\eqi G\mathfrak{M}= a^3\left(\rp{2\pi}{P_{\rm b}}\right)^2.\lb{Massa}\eqf
With \rfr{ohe} and \rfr{Massa} we can, now, consistently calculate \rfr{oge} getting
\eqi \oge = \rp{3}{(1-e^2)}\left(\rp{x_A+x_B}{s}\right)^2\left(\rp{2\pi}{P_{\rm b}}\right)^3=16.90410\ {\rm deg\ yr}^{-1}; \eqf
in this way the 1PN periastron precession is written in terms  of the four Keplerian parameters $P_{\rm b}, e, x_{\rm A}, x_{\rm B}$ and of the post-Keplerian parameter $s$.
The mismodeling in them yields
\begin{equation}\left\{\begin{array}{lll}
\delta \oge|_{x_{\rm B}}\leq 2\oge\left[\rp{\delta x_{\rm B}}{(x_{\rm A} + x_{\rm B} )}\right] = 0.01845\ {\rm deg\ yr}^{-1},\\\\
\delta \oge|_{s}\leq  2\oge\left(\rp{\delta s}{s}\right)= 0.01318\ {\rm deg\ yr}^{-1},\\\\
\delta \oge|_{x_{\rm A}}\leq  2\oge\left[\rp{\delta x_{\rm A}}{(x_{\rm A} + x_{\rm B} )}\right]= 1\times 10^{-5}\ {\rm deg\ yr}^{-1},\\\\
\delta \oge|_{\rm e}\leq  2e\oge\left[\rp{\delta e}{(1-e^2)}\right]= 2\times 10^{-6}\ {\rm deg\ yr}^{-1},\\\\
\delta \oge|_{P_{\rm b}}\leq  3\oge\left(\rp{\delta P_{\rm b}}{ P_{\rm b} }\right) = \mathcal{O}(10^{-8})\ {\rm deg\ yr}^{-1}.
 \lb{erra}
\end{array}\right.\end{equation}

Thus, the total uncertainty is
\eqi \delta\oge\leq 0.03165\ {\rm deg\ yr}^{-1},\lb{kazza}\eqf
which maps into a relative uncertainty of \eqi \rp{\delta\oge} \oge=1.8\times 10^{-3}.\lb{canc}\eqf
As a consequence, we have the important result
\eqi\Delta\dot\omega\doteq\oms-\oge=(-0.00463\pm 0.03233)\ {\rm deg\ yr}^{-1}.\lb{emho}\eqf
Every attempt to measure or constrain effects predicted by known Newtonian and post-Newtonian physics (like, e.g., the action of the quadrupole mass moment or the gravitomagnetic field), or by modified models of gravity, for the periastron of the \psr system must face with the bound of \rfr{emho}.

Should we decide to use both the post-Keplerian parameters related to the Shapiro delay  \citep{Dam86, Dam92}
\begin{equation}\left\{\begin{array}{lll}
r=T_{\odot}m_{\rm B},\\\\
s = x_{\rm A}\left(\rp{P_{\rm b}}{2\pi}\right)^{-2/3} T_{\odot}^{-1/3}M^{2/3}m_{\rm B}^{-1},
\lb{shap}
\end{array}\right.\end{equation}
for determining the sum of the masses, we would have, with \rfr{ohe},
\eqi \oge = \rp{3}{(1-e^2)}  \left( \rp{P_{\rm b}}{2\pi} \right)^{3/2} \left( \rp{r}{x_{\rm A}}  \right)^{9/4} \rp{ s^{19/4} }{(x_{\rm A}+x_{\rm B})^{5/2}},\lb{nueva}\eqf
which yields
\eqi \oge = 17.25122\pm 2.11819\ {\rm deg}\ {\rm yr}^{-1}.\eqf
The major source of uncertainty is $r$, with 2.06264 deg yr$^{-1}$; the bias due to the other parameters is about the same as in the previous case.

Let us, now, consider the second post-Newtonian contribution to the periastron precession \citep{Dam88, Wex95}
\eqi \ogge = \rp{ 3(G\mathfrak{M})^{5/2} }{c^4 a^{7/2}(1-e^2)^2}\left\{ \left[\rp {13} 2\left(\rp{m_{\rm A}^2 + m_{\rm B}^2}{M^2}\right) +\rp {32} 3\left(\rp{m_{\rm A} m_{\rm B}}{M^2}\right)\right]\right\},\eqf up to terms of order $\mathcal{O}(e^2)$.
For our system it amounts to $4\times 10^{-4}$ deg yr$^{-1}$, so that it should be taken into account in $\Delta\dot\omega$. However, it can be shown that the bias induced by the errors in $\mathfrak{M}$ and $a$ amounts to
$4\times 10^{-6}$ deg yr$^{-1}$, thus affecting the gravitomagnetic precession at the percent level.

\subsection{Summary and Discussion}
\citet{OCo04}, aware of the presence of other non-gravitomagnetic contributions to the pulsar's periastron rate,  proposed  to try to measure the  gravitomagnetic  spin-orbit precession of the orbital angular momentum \citep{Bar75a} (analogous to the Lense-Thirring node precession in the limit of a test particle orbiting a massive body) which is not affected by larger gravitoelectric contributions. However, its magnitude is $\approx (10^{-4}\ {\rm deg\ yr}^{-1})\sin\psi$, where $\psi$ is the angle between the orbital angular momentum and the pulsar's spin; thus, it would be negligible in the \psr system because of the near alignment between such vectors \citep{Sta06}, in
agreement with the observed lack of profile variations \citep{Man05,Kra06}.

In regard to the measurement of the moment of inertia of the component A via the gravitomagnetic periastron precession, our analysis has pointed out that  the bias due to the mismodelling in the 1PN gravitoelectric contribution to periastron precession--expressed in terms of the phenomenologically measured parameters $P_{\rm b}, e, x_{\rm A}, x_{\rm B}, s$--is the most important systematic error exceeding the expected gravitomagnetic rate, at present, by two orders of magnitude; the major sources of uncertainty in it are $x_{\rm B}$ and $s$, which should be measured three orders of magnitude better than now to reach the $10\%$ goal.  The projected semimajor axis $x_{\rm A}$ of A, if known one order of magnitude better than now, would induce a percent-level bias.  Instead, expressing the 1PN gravitoelectric periastron rate in terms of $P_{\rm b}, e, x_{\rm A}, x_{\rm B}, s, r$ would be definitely not competitive because the improvement required for $r$ would amount to five orders of magnitude at least. We prefer not to speculate now about the size of the improvements in timing of the \psr system which could be achieved in future.
Since the timing data of B are required as well for $x_{\rm B}$  and in view of the fact that B appears as a strong radio
source only for two intervals, each of about 10-min duration, while its pulsed emission is rather
weak or even undetectable for most of the remainder of the orbit \citep{Lyn04, Bur05}, the possibility of reaching  in a near future the required accuracy to effectively constrain $I_{\rm A}$ to $10\%$ level or better should be  considered with more skepticism than done so far. Another analysis on this topic has been recently performed by \citet{Kranew}.
\section{Testing a Uniform Cosmological Constant and the DGP Gravity with the Pulsar J0737-3039A}\lb{modified}
Since, at present, the only reason why the cosmological constant\footnote{See \citep{Cal08} and references therein for an interesting historical overview.} $\Lambda$ is believed to be nonzero relies upon the observed acceleration of the universe \citep{Rie98,Per99}, i.e. just the phenomenon for which $\Lambda$ was introduced (again), it is important to find independent observational tests of the existence of such an exotic component of the spacetime.

 Here we  put on the test the hypothesis that $\Lambda\neq 0$, where $\Lambda$ is the uniform cosmological constant of the Schwarzschild-de Sitter \citep{Stu99} (or Kottler \citep{Kot18}) spacetime, by suitably using the latest determinations of the parameters (see Table \ref{tavola}) of the double pulsar \psr\ system.

The approach followed here consists in deriving analytical expressions $\mathcal{O}_{\Lambda}$ for the effects induced by $\Lambda$ on some quantities for which empirical values $\mathcal{O}_{\rm meas}$ determined  from fitting the timing data exist.  By taking into account the known classical and relativistic effects $\mathcal{O}_{\rm known}$ affecting such quantities, the discrepancy $\Delta{\mathcal{O}}=\mathcal{O}_{\rm meas} - \mathcal{O}_{\rm known}$ is constructed and attributed to the action of $\Lambda$, which was not modelled in the pulsar data processing. Having some $\Delta\mathcal{O}$ and $\mathcal{O}_{\Lambda}$ at hand, a suitable combination $\mathcal{C}$, valid just for the case $\Lambda\neq 0$, is constructed out of  them in order to  compare $\mathcal{C}_{\rm meas}$      to $\mathcal{C}_{\rm \Lambda}$: if the hypothesis $\Lambda\neq 0$ is correct, they must be equal within the errors. Here we will use the anomalistic period $\Po$ and the periastron precession $\do
t\omega$ for which purely phenomenological determinations exist in such a way that our $\mathcal{C}$ is the ratio of $\Delta\dot\omega$ to $\Delta\Po$; as we will see, this observable is independent of $\Lambda$ but, at the same time, it retains a functional dependence on the system's parameters peculiar to the $\Lambda-$induced force and of any other Hooke-like forces.

This Section complements \citep{Ior07} in which a similar test was conducted in the Solar System by means of the latest determinations of the
secular precessions of the longitudes of the perihelia of several planets. The result by \citet{Ior07} was negative for the Schwarzschild-de Sitter spacetime with uniform $\Lambda$; as we will see, the same conclusion will be traced here in Section \ref{posec}.

A complementary approach to explain the cosmic acceleration without resorting to dark energy was followed by Dvali, Gabadadze and Porrati (DGP) in their braneworld modified model of gravity \citep{DGP}.   Among other things, it predicts effects which could be tested on a local, astronomical  scale.
In \citep{Ior07} a negative test in the Solar System was reported; as we will see in Section \ref{dgpsec},  \psr\ confirms such a negative outcome at a much more stringent level.

The overview and the conclusions are in Section \ref{conc}.
\subsection{The Impact of $\Lambda$ on the Periastron and the Orbital Period}\lb{calazza}
The Schwarzschild-de Sitter metric induces an extra-acceleration\footnote{The present test is valid for all exotic Hooke-type forces of the form $Cr$ \citep{Cal08}, with $C$ arbitrary nonzero constant.}
\citep{Rin} \eqi \mathbf{A}_{\Lambda}=\rp{1}{3}\Lambda c^2 \mathbf{r},\lb{acc}\eqf
where $c$ is the speed of light; \rfr{acc}, in view of the extreme smallness of the assumed nonzero value cosmological constant ($\Lambda \approx 10^{-52}$ m$^{-2}$), can be treated perturbatively with the standard techniques of celestial mechanics.  In \citep{Hau03} the secular precession of the pericentre of  a test particle around a central body of mass $\mathfrak{M}$ was found    to be
\eqi\dot\omega_{\Lambda} = \rp{\Lambda c^2}{2 \nkep}\sqrt{1-e^2},\lb{olam}\eqf where
\eqi\nkep \doteq \sqrt{\rp{G\mathfrak{M}}{a^3}}\eqf  is the Keplerian mean motion; $a$ and $e$ are the semimajor axis and the eccentrity, respectively, of the test particle's orbit.
Concerning a binary system, in \citep{Jet} it was shown that the equations for the relative motion are those of a test particle in a Schwarzschild-de Sitter space-time with a
source mass equal to the total mass of the two-body system, i.e. $\mathfrak{M} = \mathfrak{m}_{\rm A}+\mathfrak{m}_{\rm B}$. Thus, \rfr{olam} is valid in our case; $a$ is the semi-major axis of the
relative orbit.

Following the approach by \citet{Jet}, we will now compute $P_{\Lambda}$, i.e. the contribution of $\Lambda$ to the orbital period.
One of the six Keplerian orbital elements in terms of which it is
possible to parameterize the orbital motion in a
binary system is the mean anomaly  $\mathcal{M}$ defined as
$\mathcal{M}\doteq n(t-T_0)$, where $n$ is the mean motion and
$T_0$ is the time of pericenter passage. The mean motion $n\doteq
2\pi/ P_{\rm b}$ is inversely proportional to the time elapsed
between two consecutive crossings of the pericenter, i.e. the
anomalistic period $P_{\rm b}$. In Newtonian mechanics, for two
point-like bodies, $n$ reduces to the usual Keplerian expression
$\nkep=2\pi/\Pkep$. In
many binary systems, as in the double pulsar one, the period $P_{\rm b }$ is  accurately
determined in a phenomenological, model-independent way, so that, in principle,
it accounts for all the dynamical features of the system, not only
those coming from the Newtonian point-like terms, within the
measurement precision.

The Gauss equation for the variation of the mean anomaly, in the case of an entirely radial disturbing acceleration   $\mathcal{A}_r$
like \rfr{acc}, is \citep{roy88}
\eqi\dert{\mathcal{M}} t=n-\rp{2}{na}\mathcal{A}_r
\left(\rp{r}{a}\right)+\rp{(1-e^2)}{nae}\mathcal{A}_r\cos f,\lb{gauss}\eqf
where $f$ is the true anomaly, reckoned from the pericenter.
Using the eccentric anomaly $E$, defined as
\eqi \mathcal{M} = E - e\sin E,\eqf turns out to be more convenient.
The unperturbed Keplerian ellipse, on which the right-hand-side of \rfr{gauss} must be evaluated, is
\eqi r = a\left(1-e\cos E\right);\eqf by using
\rfr{acc} and \begin{equation}\left\{\begin{array}{lll}
\dert{\mathcal{M}}E = 1 - e\cos E,\\\\
\cos f = \rp{\cos E - e}{1 - e\cos E},
\lb{grazia}
\end{array}\right.\end{equation}
\rfr{gauss} becomes
\begin{eqnarray} & \dert E t & =  \rp{\nkep}{\left(1-e\cos E\right)}\left\{1-\rp{\Lambda c^2}{3{\nkep}^2}\left[2\left(1-e\cos E\right)^2 -\right.\right.\nonumber\\
& &-\left.\left.\rp{\left(1-e^2\right)}{e}\left(\cos E-e\right)\right]\right\}.\lb{grossa}
\end{eqnarray}
Since $\Lambda c^2/3{\nkep}^2\approx 10^{-29}$ from \rfr{grossa} it can be obtained
\begin{eqnarray} & \Po & \simeq \int_0^{2\pi}\rp{\left(1-e\cos E\right)}{\nkep}\left\{1+\rp{\Lambda c^2}{3{\nkep}^2}\left[2\left(1-e\cos E\right)^2 -\right.\right.\nonumber\\
& &-\left.\left. \rp{\left(1-e^2\right)}{e}\left(\cos E-e\right)\right]\right\}dE,
\end{eqnarray}
which integrated yields   that
\eqi\Po = \Pkep + P_{\Lambda}\eqf
with
\eqi P_{\Lambda} = \rp{\pi\Lambda c^2\left(7 + 3 e^2\right)}{3{\nkep}^3}.\lb{pla}\eqf
\subsubsection{Combining the Periastron and the Orbital Period}\lb{posec}
For the sake of convenience, from Section \ref{kazzam} let us recall the general relativistic expressions of the post-Keplerian parameters $r,s$ and $\dot\omega$;
 \begin{equation}\left\{\begin{array}{lll}
r=T_{\odot}m_{\rm B},\\\\
s = x_{\rm A}\left(\rp{P_{\rm b}}{2\pi}\right)^{-2/3} T_{\odot}^{-1/3}M^{2/3}m_{\rm B}^{-1}, \\\\
\opn = \rp{3}{\cen}\left(\Pop\right)^{-5/3}(\Ts M)^{2/3}.
\lb{shap}
\end{array}\right.\end{equation}

By means of \eqi a = \rp{c}{s}(\xa + \xb)\eqf and of the equations for $r$ and $s$ it is possible to express $\Pkep$ and $\opn$ in terms of $\Po$ and of the phenomenologically determined Keplerian and post-Keplerian parameters $\xa, \xb, r, s$ as
\begin{equation}\left\{\begin{array}{lll}
\Pkep = 2\left(\rp{2}{\Po}\right)^{1/2}\left[\pi(\xa+\xb)\right]^{3/2}\left(\rp{\xa}{r}\right)^{3/4}s^{-9/4},\\\\
\opn = \rp{3sr}{\xa(\cen)}\left(\Pop\right)^{-1}.
\lb{grazia}
\end{array}\right.\end{equation}

 In such a way we can genuinely compare them to $\Po$ and $\dot\omega$ because they do not contain quantities obtained from the third Kepler law and the general relativistic periastron precession themselves; moreover, we have expressed the sum of the masses entering both $\Pkep$ and $\opn$ in terms of $r$ and $s$, thus avoiding any possible reciprocal imprinting between the third Kepler law and the periastron rate.
At this point it is possible to construct
\eqi R \doteq \rp{\Delta\dot\omega}{\Delta P},\lb{rap}\eqf with
 \begin{equation}\left\{\begin{array}{lll}
 \Delta\dot\omega = \dot\omega-\opn,\\\\
\Delta P = \Po-\Pkep;
 \lb{rapp}
\end{array}\right.\end{equation}
note that \eqi R = R(\Po,\xa,\xb,e;\dot\omega,r,s).\eqf
By attributing   $\Delta\dot\omega$ and $\Delta P$ to the action of $\Lambda$, not modelled into the routines used to fit the \psr\ timing data, it is possible to compare $R$ to
\eqi R_{\Lambda}\doteq\rp{\dot\omega_{\Lambda}}{P_{\Lambda}}=\rp{3\sqrt{1-e^2}\Po r^{3/2}s^{9/2}}{4\pi^2(7 + 3e^2)(\xa+\xb)^3 \xa^{3/2}}\lb{Rlam}\eqf
and see if \rfr{rap} and \rfr{Rlam} are equal within the errors. Note that \rfr{Rlam} is independent of $\Lambda$ and, by definition, is able to test the hypothesis that $\Lambda\neq 0$. From Table \ref{tavola} it turns out
\eqi R_{\Lambda}=(3.4\pm 0.3)\times 10^{-8}\ {\rm s}^{-2};\eqf $R_{\Lambda}$ is a well determined quantity, different from zero at about 11 sigma level.
In regard to $R$ we have
 \begin{equation}\left\{\begin{array}{lll}
\Delta\dot\omega = -0.3\pm 2.1\ {\rm deg\ yr}^{-1},\\\\
\Delta P = 59\pm 364\ {\rm s},
\lb{rappnum}
\end{array}\right.\end{equation}
so that \eqi \left|R\right| = (0.3\pm 4)\times 10^{-11}\ {\rm s}^{-2}; \eqf $R$ is compatible with zero in such a way that its range does not overlap with the one of $R_{\Lambda}$: indeed, the upper bound on $R$ is three orders of magnitude smaller than the lower bound on $R_{\Lambda}$.
Thus, we must conclude that\footnote{In principle, also the 1PN correction to the third Kepler law \citep{Dam86} should be included in $\Delta P$, but it does not change the result.}
\eqi R \neq R_{\Lambda}.\eqf   Concerning the released uncertainties in $R$ and $R_{\Lambda}$, they must be considered as upper bounds since they have been conservatively computed by linearly adding the individual biased terms due to $\delta\Po,\delta\dot\omega,\delta e, \delta\xa,\delta\xb,\delta r,\delta s$ in order to account for the existing correlations \citep{Kra06} among  them.

The results of the present study confirm those obtained in the Solar System by taking the ratio of the estimated corrections to the standard Newtonian/Einsteinian precessions of the longitude of the perihelia $\varpi$ for different pairs of planets \citep{Ior07}.    It would be very interesting to devise analogous tests involving other observables (lensing, time delay) affected by $\Lambda$ as well recently computed in, e.g., \citep{Rug07, Ser08}.
\subsection{The Dvali-Gabadadze-Porrati Braneworld Model}\lb{dgpsec}
The approach previously outlined for $\Lambda$ can be followed also for the DGP braneworld model \citep{DGP} which recently received great attention from an observational point of view \citep{Dva03,Ior08}.

The preliminary confrontations with data so far performed refer to the perihelia of the Solar System planets. Indeed, DGP gravity predicts an extra-precession of the pericentre of a test particle   \citep{Lue03,Ior05}
\eqi \dot\omega_{\rm DGP}=\mp\rp{3}{8}\left(\rp{c}{r_0}\right)\left(1-\rp{13}{32}e^2\right),\lb{odgp}\eqf where the signs $\mp$ are related to the two different cosmological branches of the model and $r_0$ is a free-parameter set to about 5 Gpc by Type IA Supernov{\ae} data, independent of the orbit's semimajor axis. The predicted precessions of about $10^{-4}$ arcsec cy$^{-1}$ were found to be compatible with the estimated  corrections to the usual apsidal precessions of planets considered one at a time separately \citep{Ior08}, but marginally incompatible with the ratio of them for some pairs of inner planets \citep{IorAHEP}.

The effects of DGP model on the orbital period is    \citep{Ior06}
\eqi P_{\rm DGP}=\mp\rp{11}{8}\pi\left(\rp{c}{r_0}\right)\rp{a^3 (1-e^2)^2}{G\mathfrak{M}}\lb{pdgp}.\eqf

From \rfr{odgp} and \rfr{pdgp} it is possible to construct
\eqi R_{\rm DGP}\doteq\rp{\dot\omega_{\rm DGP}}{P_{\rm DGP}},\eqf which, expressed in terms of the phenomenologically determined parameters of \psr, becomes
\eqi R_{\rm DGP} = \rp{3\left(1-\rp{13}{32}e^2\right)\Po r^{3/2} s^{9/2}}{22\pi (1-e^2)\left(\xa + \xb\right)^3 \xa^{3/2}}.\lb{Rdgp}\eqf
Putting the figures of Table \rfr{tavola} into \rfr{Rdgp} and computing the uncertainty as done in the case of $\Lambda$ yields
\eqi R_{\rm DGP} =  (1.4\pm 0.1)\times 10^{-7}\ {\rm s}^{-2}.\eqf
As can be noted, the lower bound of $R_{\rm DGP}$ is four orders of magnitude larger than the upper bound of $R$, so that we must conclude that, also in this case,
\eqi R\neq R_{\rm DGP}.\eqf
The outcome by \citet{IorAHEP} is, thus, confirmed at a much more stringent level.

An analysis of type Ia supernov{\ae} (SNe Ia) data disfavoring  DGP model can be found in \citep{Ben05}.
\subsection{Summary and Discussion}\lb{conc}
In this Section we used the most recent determinations of the orbital parameters of the double pulsar binary system \psr\ to perform local tests of  two complementary approaches to the issue of the observed acceleration of the universe: the uniform cosmological constant $\Lambda$ in the framework of the known general relativistic laws of gravity  and the multidimensional braneworld model by Dvali, Gabadadze and Porrati which, instead, resorts to a modification of the currently known laws of gravity. Since, at present, there are no observational evidences for such theoretical schemes other than just the cosmological phenomenon for which they were introduced, it is important to put them on the test independently of the cosmological acceleration itself.
It is worthwhile noting that the results for $\Lambda$ hold also for any other Hooke-like additional force proportional to $r$.

To this aim, we  considered the phenomenologically determined  the periastron precession $\dot\omega$ and the orbital period $\Po$ of \psr\ by contrasting them to the predicted  1PN periastron rate $\opn$ and the Keplerian period $\Pkep$. With such discrepancies we constructed the ratio $R\doteq\Delta\omega/\Delta P$  by finding it compatible with zero: $|R| = (0.3\pm 4) \times 10^{-11} \ {\rm s}^{-2}$.  Then, we compared $R$ to the predicted ratios for the effects of $\Lambda$ and the DGP gravity-not modeled in the pulsar data processing-on the periastron rate and the orbital period by finding $R_{\Lambda} = (3.4\pm 0.3) \times 10^{-8} \ {\rm s}^{-2}$ and $R_{\rm DGP} = (1.4\pm 0.1) \times 10^{-7} \ {\rm s}^{-2}$, respectively. Thus, the outcome of such a local test is neatly negative, in agreement with other local tests recently performed in the Solar System by taking the ratio of the non-Newtonian/Einsteinian rates of the perihelia for several pairs of planets.

\section{The 1PN Secular Effects on the Mean Anomaly in Binary Pulsar Systems}\lb{meananomaly}
According to the Einstein's General Theory of Relativity (GTR),
the post-Newtonian gravitoelectric two-body acceleration of order
$\mathcal{O}(c^{-2})$ (1PN) is, in the post-Newtonian centre of
mass frame (see \citep{Dam85} and, e.g., \citep{port04} and references therein)\eqi
{\bds a}^{\rm GE}=\rp{G\mathfrak{M}}{c^2
r^3}\left\{\left[\rp{G\mathfrak{M}}{r}(4+2\nu)-(1+3\nu)v^2+\rp{3\nu}{2r^2}(\bds
r\cdot \bds v )^2\right]\bds r+(\bds r\cdot\bds v)(4-2\nu)\bds v
\right\},\lb{acc}\eqf where $\bds r$ and $\bds v$ are the relative
position and velocity vectors, respectively, $G$ is the Newtonian
constant of gravitation, $c$ is the speed of light, $\mathfrak{m}_1$ and $\mathfrak{m}_2$
are the rest masses of the two bodies, $\mathfrak{M}\doteq \mathfrak{m}_1+\mathfrak{m}_2$ and $\nu\doteq m_1
m_2/m^2<1.$

Let us recall from Section \ref{calazza} that the orbital phase can be characterized by the mean anomaly
$\mathcal{M}$ defined as
\eqi\mathcal{M}\doteq n(t-T_0)\lb{meananom},\eqf where the
unperturbed mean motion $n$ is defined as \eqi
n\doteq\rp{2\pi}{P_b}.\eqf In it $P_b$ is the anomalistic period, i.e. the
time elapsed between two consecutive pericentre crossings, which
is $2\pi\sqrt{a^3/G\mathfrak{M}}$ for an unperturbed Keplerian ellipse, and
$T_0$ is the date of a chosen pericentre passage.

The variation of the mean anomaly can be written, in general, as\footnote{See also \citet{Linsen}.}
\eqi \dert{\mathcal{M}}{t}=n-2\pi\left(\rp{\dot
P_b}{P_b^2}\right)(t-T_0)-\rp{2\pi}{P_b}\left(\dert{T_0}{t}\right).\lb{general}\eqf
 The second term of the right-hand side of
\rfr{general} accounts for any possible variation of the
anomalistic period. The third term, induced by any small
perturbing acceleration with respect to the Newtonian monopole,
whether relativistic or not, is the
change of the time of the pericentre passage, which we will define as
\eqi\dert{\xi}{t}\doteq -\rp{2\pi}{P_b}\left(\dert{T_0}{t}\right).\eqf It can be
calculated with the aid of the Gauss\footnote{For a different approach based on a modified form of the Lagrange planetary equations see \citep{Cal97}.}  perturbative equation
\citep{roy88} \eqi\dert{\xi}{t}=
-\rp{2}{na}\mathcal{A}_r\left(\rp{r}{a}\right)-\sqrt{1-e^2}\left(\dert{\omega}{t}+\cos
i\dert{\Omega}{t}\right),\lb{manom} \eqf where  $i,\Omega$  are  the
inclination and the longitude of the ascending node, respectively, of the orbit. In order to obtain the
secular effects, we must evaluate  the right-hand-side of
\rfr{manom} on the unperturbed Keplerian ellipse and, then,
average the result over one orbital revolution.

%

We will now consider \rfr{acc} as perturbing acceleration.
 Let us start with the first term of the right-hand-side of
\rfr{manom}. By defining \eqi\left\{
\begin{array}{lll}\lb{defz}
A&\doteq &\rp{(G\mathfrak{M})^2}{c^2}(4+2\nu),\\\\
B&\doteq &-\rp{G\mathfrak{M}}{c^2}(1+3\nu),\\\\
C&\doteq &\rp{G\mathfrak{M}}{c^2}\left(4-\rp{\nu}{2}\right),
\end{array}
\right. \eqf it is possible to obtain from \rfr{acc}\eqi \mathcal{A}_r^{\rm GE}=\rp{A}{r^3}+B\left(\rp{v^2}{r^2}\right)+C\left(\rp{\dot r}{r^2}\right).\lb{radial}\eqf
Now the term $-2 \mathcal{A}_r r/na^2$, with $\mathcal{A}_r$ given by \rfr{radial}, must
be evaluated on the unperturbed Keplerian ellipse characterized by
\eqi\left\{
\begin{array}{lll}\lb{kepl}
r&=&\rp{a(1-e^2)}{1+e\cos f},\\\\
\dot r&=&\rp{nae\sin f}{\sqrt{1-e^2}},\\\\
v^2&=&\rp{n^2 a^2}{(1-e^2)}(1+e^2+2e\cos f)
\end{array}
\right. \eqf where $f$ is the true anomaly, and averaged over one
orbital period by means of
\begin{equation}
\rp{d t}{P_b}=\rp{r^2 df}{2\pi a^2\sqrt{1-e^2}}.
\end{equation}
Thus, \eqi -\left(\rp{2}{na}\mathcal{A}_r^{\rm GE}\rp{r}{a}\right)\left(\rp{dt}{P_b}\right)=-\rp{1}{na^4\pi\sqrt{1-e^2}}\left[A+Brv^2+Cr(\dot
r )^2\right]df .\lb{ert}\eqf In the expansion of $r$ in \rfr{ert}
the terms of order $\mathcal{O}(e^4)$ are retained. The final
result is \eqi\left\langle -\rp{2}{na}\mathcal{A}_r^{\rm GE}\rp{r}{a}\right\rangle_{P_b}= \rp{nG\mathfrak{M}}{c^2
a\sqrt{1-e^2}}H(e;\nu),\lb{inter}\eqf with \eqia H\simeq
&-&2(4+2\nu)+(1+3\nu)\left(2+e^2+\rp{e^4}{4}+\rp{e^6}{8}\right)-\nonumber\\
&-&\left(4-\rp{\nu}{2}\right)\left(e^2 +\rp{e^4}{4}+\rp{e^6}{8}
\right).\eqfa

The post-Newtonian gravitoelectric secular rate of pericentre is
independent of $\nu$ and is given by the well known formula
\eqi\left.\dert{\omega}{t}\right|_{\rm GE }=\rp{3nG\mathfrak{M}}{c^2
a(1-e^2)}\lb{perge},\eqf while there are no secular effects on the
node.

The final expression for the post-Newtonian secular rate of the
mean anomaly is obtained by combining \rfrs{inter}{perge} and by
considering that, for a two-body system, it is customarily to
write
\eqi\rp{nG\mathfrak{M}}{c^2}=\left(\rp{P_b}{2\pi}\right)^{-5/3}(T_{\odot}M)^{2/3}.\eqf
 It is
\eqi\left.\dert{\mathcal{\xi}}{t}\right|_{\rm GE}=-9\left(\rp{P_b}{2\pi}\right)^{-5/3}(T_{\odot}M)^{2/3}(1-e^2)^{-1/2}F(e;\nu)\lb{fine}\eqf
with\eqi
F=\left[\left(1+\rp{e^2}{3}+\rp{e^4}{12}+\rp{e^6}{24}+...\right)-\rp{2}{9}\nu\left(
1+\rp{7}{4}e^2+\rp{7}{16}e^4+\rp{7}{32}e^6+...\right)\right].\lb{finee}\eqf
Note that \rfr{fine} is negative because \rfr{finee} is always
positive; thus the crossing of the apsidal line occurs at a later
time with respect to the Kepler-Newton case.

Note that, for $\nu\rightarrow 0$, i.e. $m_1\ll m_2$, \rfr{fine}
does not vanish and, for small eccentricities, it becomes
\eqi\left.\dert{\mathcal{\xi}}{t}\right|_{\rm GE}\simeq
-\rp{9nG\mathfrak{m}_2}{c^2
a\sqrt{1-e^2}}\left(1+\rp{e^2}{3}\right),\lb{myles}\eqf which
could be used for planetary motion in the Solar System \citep{Ior04}.
E.g., for Mercury it yields a secular effect of almost $-130$ arcsec
cy$^{-1}$. It is important to note that the validity of the
present calculations has also been numerically checked by
integrating over 200 years the Jet Propulsion Laboratory (JPL)
equations of motion of all the planets of the Solar System with
and without the gravitoelectric $1/c^2$ terms in the dynamical
force models \citep{esta} in order to single out just the
post-Newtonian gravitoelectric effects. They fully agree with
\rfr{myles} (E.M. Standish, private communication, 2004). Another
analytical calculation of the post-Newtonian general relativistic
gravitoelectric secular rate of the mean anomaly was performed
\citep{rubincam 1977} in the framework of the Lagrangian perturbative
scheme for a central body of mass $\mathfrak{M}$-test particle system. \citet{rubincam 1977} starts from the space-time line
element of the Schwarzschild metric written in terms of the
Schwarzschild radial coordinate $r^{'}$. Instead, \rfr{acc} and
the equations of motion adopted in the practical planetary data
reduction at, e.g. JPL, are written in terms of the standard
isotropic radial coordinate $r$ related to the Schwarzschild
coordinate by $r^{'}=r(1+G\mathfrak{M}/2c^2 r)^2$. As a consequence, the
obtained exact expression \eqi \left.\dert{\xi}{t}\right|_{\rm
GE }^{(\rm Rubincam )}=\rp{3nG\mathfrak{M}}{c^2a\sqrt{1-e^2}},\lb{rubi}\eqf contrary
to the pericentre case, agrees neither with \rfr{myles} nor with
the JPL numerical integrations yielding, e.g., a secular advance
of $+42$ arcsec cy$^{-1}$ for Mercury.  For a better understanding of such comparisons, let us note that both the numerical analysis by Standish and
\citet{rubincam 1977} are based on the $\dot P_{\rm b}=0$ case; $n$ gets canceled by construction in the Standish calculation, while in \citep{rubincam 1977} the numbers are put just into \rfr{rubi}, which is the focus of that work.

\subsection{Testing Gravitational Theories with Binary Pulsars}
In general, in the pulsar's timing data reduction
process\footnote{For all general aspects of the binary pulsar
systems see \citep{wex01, sta03} and references therein.} five
Keplerian orbital parameters and a certain number of
post-Keplerian parameters are determined with great accuracy in a
phenomenological way, independently of any gravitational theory
\citep{wex01, sta03}.  The Keplerian parameters are the projected
semimajor axis $x=a\sin i/c$, where $i$ is the angle between the
plane of the sky, which is normal to the line of sight and is
assumed as reference plane, and the pulsar's orbital plane, the
eccentricity $e$, the orbital period $P_b$, the time of periastron
passage $T_0$ and the argument of periastron $\omega_0$ at the
reference time $T_0$. The most commonly used post-Keplerian
parameters are the periastron secular advance $\dot\omega$, the
combined time dilation and gravitational redshift due to the
pulsar's orbit $\gamma$, the variation of the anomalistic period
$\dot P_b$, the range $r$ and the shape $s$ of the Shapiro delay.
These post-Keplerian parameters are included in the timing models
\citep{wex01, sta03} of the so called Roemer, Einstein and Shapiro
$\Delta_{\rm R}, \Delta_{\rm E},\Delta_{\rm S}$
delays\footnote{For the complete expression of the timing models
including, e.g., also the delays occurring in the Solar System due
to the solar gravity see \citep{wex01, sta03}.} occurring in the
binary pulsar system\footnote{The aberration parameters $\delta_r$
and $\delta_{\theta}$ are not, in general, separately measurable.}
\eqi \left\{
\begin{array}{lll}\lb{delays}
\Delta_{\rm R}&=& x\sin\omega[\cos
E-e(1+\delta_r)]+x\cos\omega\sin
E\sqrt{1-e^2(1+\delta_{\theta})^2},\\\\
\Delta_{\rm E}&=&\gamma\sin E,\\\\
\Delta_{\rm S}&=&-2r\ln\{1-e\cos E-s[\sin\omega(\cos
E-e)+\sqrt{1-e^2}\cos\omega\sin E]\},
\end{array}
\right. \eqf where $E$ is the eccentric anomaly defined
as
$E-e\sin E=\mathcal{M}$. $\cos E$ and $\sin E$ appearing in
\rfr{delays} can be expressed in terms of $\mathcal{M}$ by means
of the following elliptic expansions \citep{vinti} \eqi \left\{
\begin{array}{lll}\lb{dalam}
\cos E&=&-\rp{e}{2}+\sum_{j=1}^{\infty}\rp{2}{j^2}\rp{d}{de}[J_j (je)]\cos (j\mathcal{M}),\\\\\
\sin E&=&\rp{2}{e}\sum_{j=1}^{\infty}\rp{J_j(je)}{j}\sin
(j\mathcal{M}),
\end{array}
\right. \eqf where $J_j(y)$ are the Bessel functions defined as
\eqi \pi J_j(y)=\int_0^{\pi}\cos(j \theta-y\sin\theta)d\theta.\eqf
The relativistic secular advance of the mean anomaly \rfr{fine}
can be accounted for in the pulsar timing modelling by means of
\rfr{dalam}.


In a given theory of gravity, the post-Keplerian parameters can be
written in terms of the mass of the pulsar $\mathfrak{m}_p$ and of the
companion $\mathfrak{m}_c$. In general,  $\mathfrak{m}_p$ and $\mathfrak{m}_c$ are unknown; this
means that the measurement of only one post-Keplerian parameter,
say, the periastron advance, cannot be considered as a test of a
given theory of gravity because one would not have a theoretically
calculated value to be compared with the phenomenologically
measured one.  In GTR the previously quoted post-Keplerian
parameters are \citep{Dam86} \eqi \left\{
\begin{array}{lll}
\dot\omega &=&
3\left(\rp{P_b}{2\pi}\right)^{-5/3}(T_{\odot}{M})^{2/3}(1-e^2)^{-1},\\\\
\gamma &=&
e\left(\rp{P_b}{2\pi}\right)^{1/3}T_{\odot}^{2/3}{M}^{-4/3}m_c(m_p+2m_c),\\\\
\dot P_b
&=&-\rp{192\pi}{5}T_{\odot}^{5/3}\left(\rp{P_b}{2\pi}\right)^{-5/3}\rp{
\left(1+\rp{73}{24}e^2+\rp{37}{96}e^4\right)}{(1-e^2)^{7/2}}\rp{m_p
m_c }{{M}^{1/3}},\label{dpdt}\\\\
r&=&T_{\odot}m_c,\\\\
s&=&xT_{\odot}^{-1/3}\left(\rp{P_b}{2\pi}\right)^{-2/3}\rp{{M}^{2/3}}{m_c}
\end{array}
\right. \eqf It is important to note that the relativistic
expression of $\dot P_b$ in \rfr{dpdt}, should not be confused
with $\left.\dot\xi\right|_{\rm GE}$ of  \rfr{fine}. Indeed, it
refers to the shrinking of the orbit due to gravitational wave
emission which vanishes in the limit $\nu\rightarrow 0$, contrary
to \rfr{fine} which expresses a different, independent phenomenon.
The measurement of two post-Keplerian orbital parameters allows to
determine $\mathfrak{m}_p$ and $\mathfrak{m}_c$, assumed the validity of a given theory
of gravity\footnote{This would still not be a test of the GTR
because the masses must be the same for all the theories of
gravity, of course.  }. Such values can, then, be inserted in the
analytical expressions of the remaining post-Keplerian parameters.
If the so obtained values are equal to the measured ones, or the
curves for the $2+N$, with $N\geq 1$, measured post-Keplerian
parameters in the $\mathfrak{m}_p-\mathfrak{m}_c$ plane all intersect in a well
determined $(\mathfrak{m}_p,\mathfrak{m}_c)$ point, the theory of gravity adopted is
consistent. So, in order to use the pulsar binary systems as
valuable tools for testing GTR the measurement of at least three
post-Keplerian parameters is required. The number of
post-Keplerian parameters which can effectively be determined
depends on the characteristics of the particular binary system
under consideration. For the pulsar-neutron star PSR B1913+16
system \citep{hul75} the three post-Keplerian parameters
$\dot\omega,\gamma$ and $\dot P_b$ were measured with great
accuracy. For the pulsar-neutron star PSR B1534+12 system
\citep{sta02} the post-Keplerian parameters reliably measured are
$\dot\omega,\gamma, r$ and $s$. For the
pulsar-pulsar PSR J0737-3039 A+B system  the same four
post-Keplerian parameters as for PSR B1534+12 are available plus
$\dot P_b$ and a further constraint on $m_p/m_c$ coming from the
measurement of both the projected semimajor axes. On the contrary,
in the pulsar-white dwarf binary systems, which are the majority
of the binary systems with one pulsar and present almost circular
orbits, it is often impossible to measure $\dot\omega$ and
$\gamma$. Up to now, only $r$ and $s$ have been measured, with a
certain accuracy, in the PSR B1855+09 system \citep{kas94}, so that
it is impossible to use its data for testing the GTR as previously
outlined.
\subsection{The Secular Decrease of the Mean Anomaly and the Binary Pulsars}
%
Let us investigate the magnitude of the mean anomaly precession in
some systems including one or two radiopulsars.

For PSR B1913+16 we have \citep{weistay05} $m_p=1.4414$, $m_c=1.3867,e=0.6171338,$
$P_b=0.322997448930$ d. Then, $\nu=0.2499064$, $F=1.04459537192$
and $\left.\dot\xi\right|_{\rm GE }=-10.422159$ deg yr$^{-1}$.
 For PSR
J0737-3039 A we have  $\nu=0.249721953643$, $F=0.946329857430$. Thus,
$\left.\dot\xi\right|_{\rm GE }=-47.79$ deg yr$^{-1}$. This
implies that the ratio of the post-Newtonian gravitoelectric
secular rate of the mean anomaly to the mean motion amounts to
$\sim 10^{-5}$. Let us see if such post-Newtonian shift is
detectable from quadratic fits of the orbital phases of the form
$\mathcal{M}=a_0+b_0t+c_0t^2$. For PSR B1913+16 the quadratic
advance due to the gravitational wave emission over thirty years
amounts to ($\dot P_b=-2.4184\times 10^{-12} $)\eqi \Delta {
\mathcal{M} }=-\pi\left(\rp{\left.\dot P_b\right|_{\rm gw
}}{P^2_b}\right)(t-T_0)^2=0.5\ {\rm deg},\eqf with an uncertainty
$\delta(\Delta { \mathcal{M} } )$ fixed to $0.0002$ deg by
$\delta\dot P_b=0.0009\times 10^{-12}$. The linear shift due to
\rfr{fine}  amounts to
\eqi\Delta{\mathcal{M}}=\left.\dot\xi\right|_{\rm GE
}(t-T_0)=-312.6647\ {\rm deg}\eqf over the same time interval. The
uncertainty in\footnote{The sum of the masses and the semimajor axis entering $n$ are determined from timing data processing independently of $\mathcal{M}$ itself, e.g. from the periastron rate and the projected barycentric semimajor axis.} $n$ amounts to $1\times 10^{-9}$ deg yr$^{-1}$ due
to $\delta P_b=4\times 10^{-12}$ d. For PSR J0737-3039 A
 the gravitational wave emission over three years
($\dot P_b=-1.20\times 10^{-12}$) induces a quadratic shift of
$0.008$ deg, with an uncertainty $\delta(\Delta { \mathcal{M} } )$
fixed to $0.0005$ deg by $\delta\dot P_b=0.08\times 10^{-12}$. The
linear shift due to \rfr{fine} amounts to -143.3700 deg over the
same time interval. The uncertainty in $n$ amounts to $7\times
10^{-7}$ deg yr$^{-1}$ due to $\delta P_b=2\times 10^{-10}$ d.
Thus,  it should be possible to extract
$\left.\dot\xi\right|_{\rm GE}$ from the measured coefficient
$b_0$; both the corrupting bias due to the uncertainties in the
quadratic signature and the errors in $n$ would be negligible.

Measuring $\left.\dot\xi\right|_{\rm GE}$ as a further
post-Keplerian parameter would be very useful in those scenarios
in which some of the traditional post-Keplerian parameters are
known with a modest precision or, for some reasons, cannot be
considered entirely reliable\footnote{The measured value of the
derivative of the orbital period $\dot P_b$ is aliased by several
external contributions which often limit the precision of the
tests of competing theories of gravity based on this
post-Keplerian parameter \citep{sta03}. }. E.g., in the double pulsar
system PSR J0737-3039 A+B the parameters $r$ and $\gamma$ are
measured with a relatively low accuracy \citep{Lyn04}. Moreover,
there are also pulsar binary systems in which only the periastron
rate has been measured \citep{kas99}: in this case the knowledge of
another post-Keplerian parameter would allow to determine the
masses of the system, although it would not be possible to
constraint alternative theories of gravity.

\subsection{Overview and Discussion}
In this Section we have analytically derived for a
two-body system in eccentric orbits the secular variation
$\left.\dot\xi\right|_{\rm GE}$ yielding the  post-Newtonian
general relativistic gravitoelectric part of the precession of the mean anomaly not due to the variation of the orbital period.
For a complementary analysis, see \citet{Linsen}.
In the limit of small
eccentricities and taking the mass of one of the two bodies
negligible, our results have been compared to the outcome of a
numerical integration of the post-Newtonian general relativistic
gravitoelectric equations of motion of the planets of the Solar
System performed by JPL: the agreement between the analytical and
numerical calculation is complete. Subsequently, we have
investigated the possibility of applying the obtained results to
the binary systems in which one pulsar is present. In particular,
it has been shown that the variation of the orbital period
$\left.\dot P_b\right|_{\rm gw}$ due to gravitational wave
emission and the effect derived by us are different ones. Indeed,
the post-Newtonian gravitoelectric precession of the mean anomaly,
which is always negative, is related to the secular increase of
the time of pericentre passage and occurs even if the orbital
period does not change in time. A quadratic fit of the orbital
phase of the pulsar would allow to measure
$\left.\dot\xi\right|_{\rm GE}$ because the biases due to the
errors in the quadratic shift due to $\dot P_b$ and in the linear
shift due of the mean motion $n$ are smaller. The use of
$\left.\dot\xi\right|_{\rm GE}$ as a further post-Keplerian
parameter would allow to improve and enhance the tests of
post-Newtonian gravity especially for those systems in which only
few post-Keplerian parameters can be reliably measured.


\end{document}